\documentclass[superscriptaddress,aps,prl]{revtex4}
\usepackage{color}
\usepackage{amsmath}    
\usepackage{amssymb}
\usepackage{graphicx}   
\usepackage{subfigure}
\usepackage{bm}

\begin{document}     

\title{Topological dimension tunes activity patterns in hierarchical modular network models}  \author{Ali Safari} \author{Paolo Moretti} 
\affiliation{Institute for Materials Simulation, Friedrich-Alexander-Universit\"at Erlangen-N\"urnberg, Dr.-Mack-Str. 77 90762 F\"urth, Germany}
\author{Miguel A. Mu\~noz}
\affiliation{Departamento de Electromagnetismo y F\'isica de la Materia e Instituto Carlos I de F\'isica Te\'orica y Computacional,
Universidad de Granada, Granada E-18071, Spain}

\begin{abstract} Connectivity patterns of relevance in neuroscience and systems biology can be encoded in hierarchical modular networks (HMNs). Moreover, recent studies highlight the role of hierarchical modular organization in shaping brain activity patterns, providing an excellent substrate to promote both the segregation and integration of neural information. Here we propose an extensive numerical analysis of the critical spreading rate (or ``epidemic'' threshold) --separating a phase with endemic persistent activity from one in which activity ceases-- on diverse HMNs.  By employing analytical and computational techniques we determine the nature of such a threshold and scrutinize how it depends on general structural features of the underlying HMN.  We critically discuss the extent to which current graph-spectral methods can be applied to predict the onset of spreading in HMNs, and we propose the network topological dimension as a relevant and unifying structural parameter, controlling the epidemic threshold.  \end{abstract}

\maketitle
\section{Introduction}
A surprising variety of biosystems display hierarchical modular architectures, an incomplete list of which includes the human brain structural network, or connectome \cite{Sporns2005,Hagmann2008}, metabolic and regulatory networks \cite{Jeong2000,Ravasz2002} and fiber networks in connective tissue \cite{Gautieri2011}. All such examples point to a common conjecture: hierarchical modular organization is capable of steering function and activity into localized patterns that provide an excellent tradeoff between segregation of functions (located at diverse moduli) and their integration across very diverse scales, constituting an elegant solution of the so-called segregation-integration dilemma \cite{Tononi1994,Arenas2006,Zhou2006,MuellerLinow2008,Bullmore2012,Sporns2013segreinte,Diez2015}. Further examples of such remarkable dynamic signatures in fields beyond the biological realm have been recently pointed out for parallel processing \cite{Sollich2014}, information retrieval \cite{Agliari2015_PRL}, as well as random and quantum walks \cite{Agliari2016}.

From the perspective of modeling techniques, hierarchical modular organization is often encoded in simple mathematical models of synthetic hierarchical modular networks (HMNs) \cite{Kaiser2007}. Song {\it et al.} first noted that in HMNs the mutual distance between highly connected hubs is higher than in scale-free (SF) networks \cite{Song2006}. As a consequence, hubs and their neighborhoods are not easily exposed to functional overloads; even if load increases in some location, it mostly remains localized in a subnetwork and does not greatly affect activity at other hubs. Song {\it et al.} \cite{Song2006} proposed that the hierarchical (or fractal) organization of functional modules (e.g. metabolic networks) is an evolutionary constraint imposed by the need for network robustness. The concept of increased distances and path lengths in HMNs --as opposed to standard scale-free and small-world models-- was later formalized through the observation that HMNs are endowed with a finite topological dimension, $D$, by Gallos {\it et al.} \cite{Gallos2012}. Let us recall that $D$ quantifies how the number $N_r$ of nodes in the local neighborhood of an arbitrary node grows with the distance, $r$ from it, $N_r\sim r^D$): lower values of $D$ amount to higher distances between network hotspots, while, formally, networks with the small-world property have $D \rightarrow \infty$.

The use of simple dynamical models have proven effective as a probing tool to understand the propagation of information or ``activity'' in biological networks.  For instance, paramount features of brain activity have been first highlighted borrowing models from quantitative epidemiology \cite{Kaiser2007,Moretti2013}; in the context of activity spreading in cortical networks or {\it connectomes} --which are well represented by HMN architectures-- it has been recently shown that simple ``epidemic'' processes of activity propagation such as susceptible-infected-susceptible (SIS) model are very convenient \cite{Haimovici,Moretti2013}. In SIS dynamics, nodes can be either active (infected I) or inactive (susceptible S); in terms of neural dynamics an active node corresponds to an active region in the brain, which can activate an inactive neighboring region with a given probability per unit time $\lambda$, and which are deactivated at rate $\mu$ (which we set to $1$ without loss of generality) due to exhaustion of synaptic resources. In the standard critical point scenario, the steady-state average fraction of active nodes $\rho^\infty$ (or prevalence) is zero for low $\lambda$ and non-zero for high $\lambda$, these two regimes being separated by a critical value $\lambda=\lambda_c$, at which scale invariant dynamic patterns are recorded. In HMNs instead, there is a whole range of $\lambda$ values where scale-invariance is observed, i.e. a Griffiths phase \cite{Moretti2013,Villa2015}. The origin of such {\it generic} (i.e. occurring in an extended region in parameter space) scale-invariant behavior is believed to be mostly structural: hierarchical modular organization promotes the emergence of rare regions where activity tends to remain localized for large times even if finally it becomes extinct owing to fluctuations.

 This is just an example of a general mechanism by which structural disorder can induce Griffiths phases, with generic critical-like scale-invariant features in complex networks \cite{Vojta2006,Munoz2010}. The structural origin of critical-like behavior is in agreement with renormalization arguments, which show how in hierarchical networks even standard percolation produces generic power-law distributions of connected component sizes, a feature otherwise normally ascribed to the critical point (or percolation threshold) \cite{Boettcher2009,Friedman2013}.

Here, we propose the numerical study of the onset of spreading --the epidemic threshold-- for the SIS in synthetic HMNs, and its relationship --employing spectral graph analyses-- with structural parameters controlling the HMN architecture. We will highlight the topological dimension, $D$, as the relevant feature that can tune and control the value of the epidemic threshold $\lambda_c$.  

\begin{figure} \includegraphics[scale=.4]{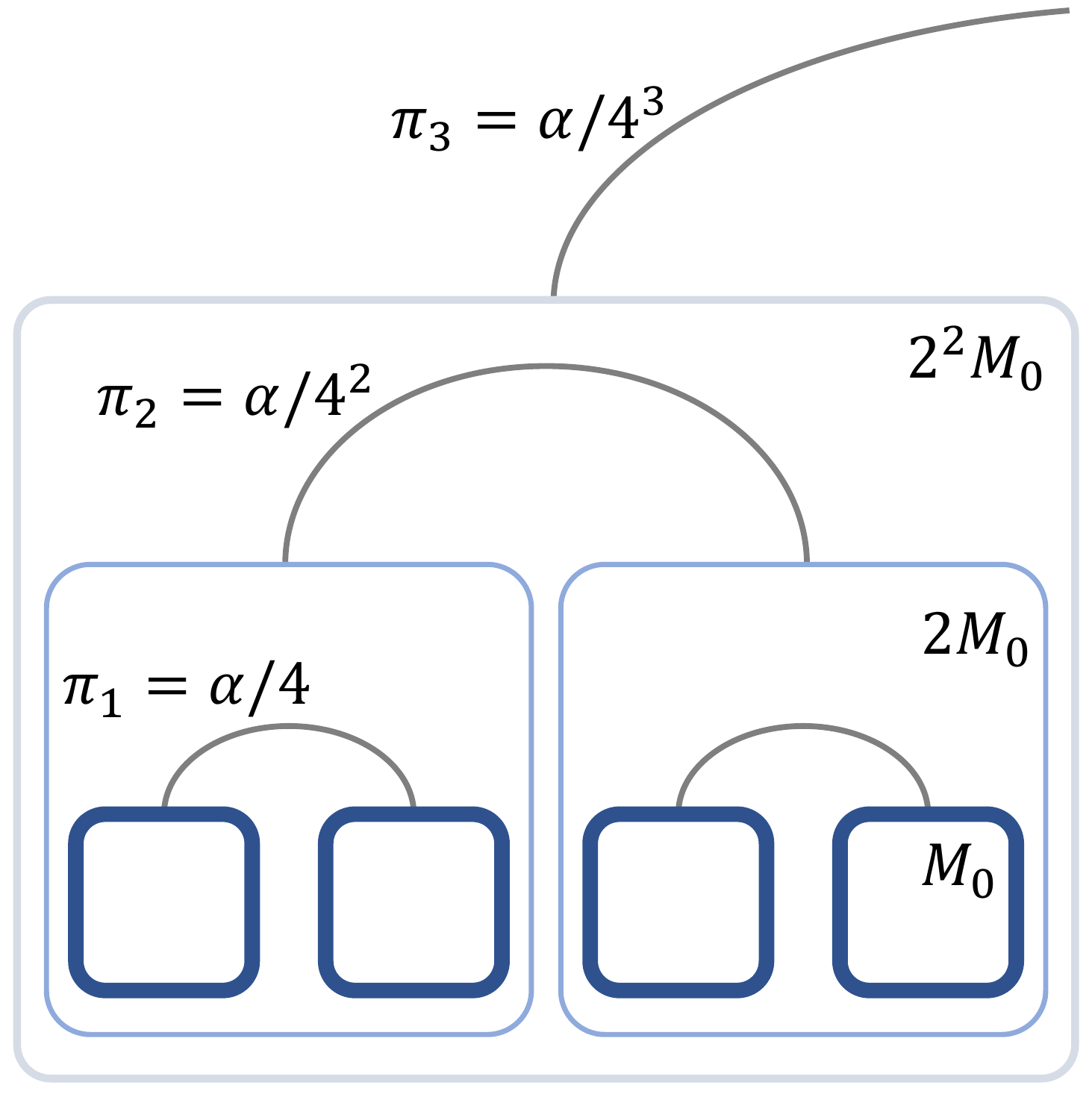} \caption{Diagram of the construction method adopted to generate synthetic hierarchical modular networks (HMNs). Densely connected modules of size $M_0$ represent the seed of the network structure (fully connected here). At each level $i$, modules of size $M_i=2^iM_0$ is formed as follows: pairs of modules of size $M_{i-1}$ are linked, by establishing random connections between their constituent nodes, with probability $\pi_i=\alpha/4^i$, where the connectivity strength, $\alpha$, is a key parameter controlling the  resulting architecture.  }\label{fig:diagram_hmn} \end{figure}

\section{Hierarchical modular networks (HMNs)}
We recall the general definition of a HMN, as a network in which smaller, more densely connected modules are clustered into larger and less densely connected super-modules, in an iterative fashion that spans several scales, or hierarchical levels. Diverse algorithms to generate synthetic HMNs have been proposed in the literature \cite{Kaiser2007,Wang2011,Moretti2013,Odor2015}. Here, we consider the model proposed in \cite{Moretti2013} motivated by previous works on optimal HMN architectures \cite{Kaiser2010}. Networks of this type will be undirected and unweighted, although generalizations to the weighted and directed cases can be introduced straightforwardly.

A schematic description of this method is shown in Fig. \ref{fig:diagram_hmn}. We call $\alpha$ the {\it connectivity strength} of a HMN, as it will appear as a chief parameter controlling the emerging topology.  The network is organized in densely connected modules of size $M_0$, which represent the level $0$ of a hierarchy of links. At each hierarchical level $i>0$, super-modules of size $M_i=2^iM_0$ are formed, joining sub-modules of size $M_{i-1}$ by wiring their respective nodes with probability $\pi_i=\alpha /4^{i}$: the average number of links between two modules at level $i$ will thus be $n_i=\pi_iM_{i-1}^2=\alpha (M_0/2)^2$, i.e. proportional to $\alpha$, regardless of the value of $i$.

\begin{figure}
\includegraphics[scale=.6]{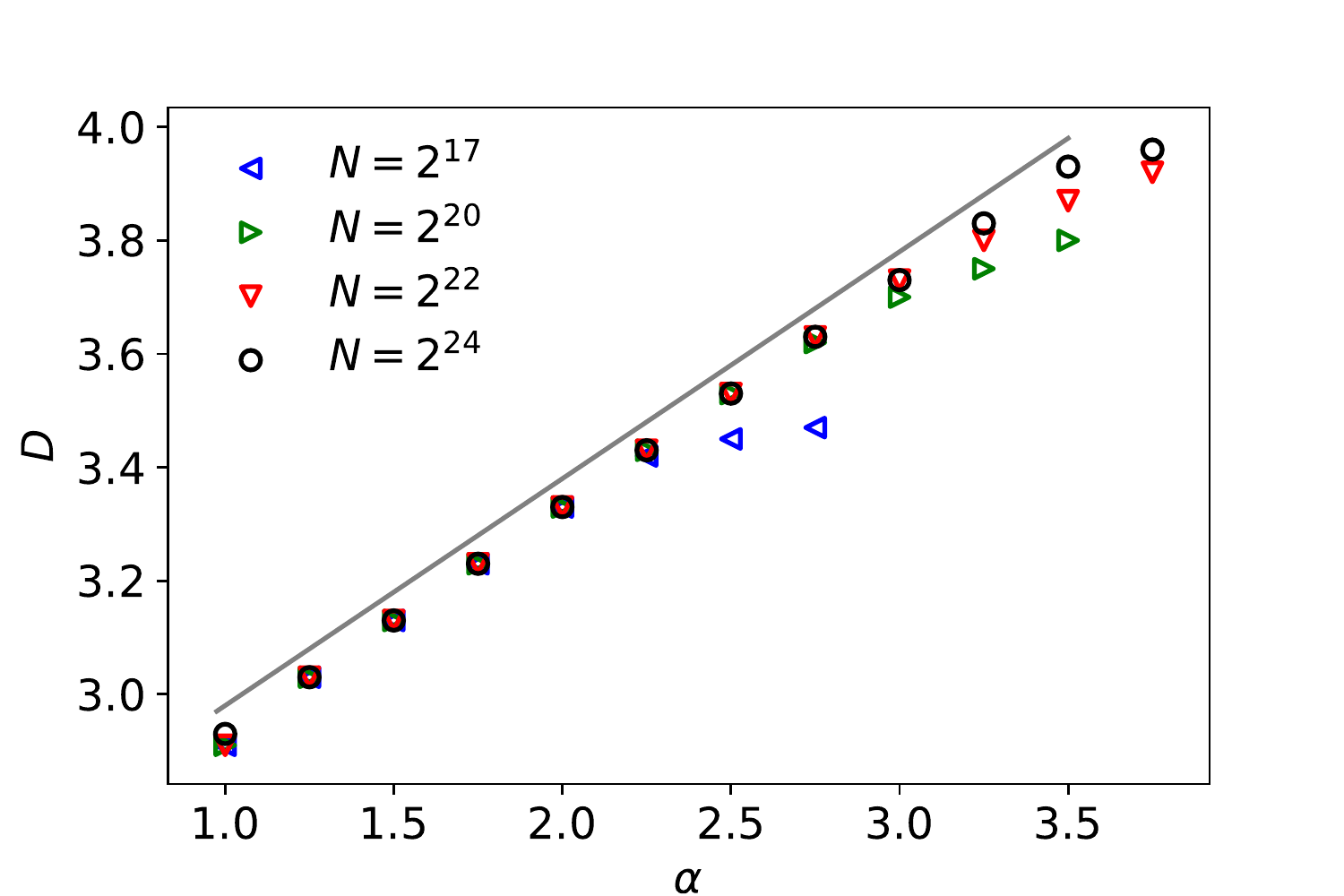}
\caption{Dependence of the topological dimension $D$ of a HMN on the connectivity strength $\alpha$. $D$ increases monotonically with $\alpha$, converging to a linear dependence for large system sizes. HMNs are considered, with $M_0=4$ and thus $n_i=4\alpha$.}\label{fig:topo}
\end{figure}

It was shown that this construction method ensures the scalability of the network structure, so that the average degree and the topological dimension $D$ reach asymptotic values for large $N$ \cite{Moretti2013}. In the $N\to \infty$ limit, the effect of lowest-level modules becomes negligible (relegated to transient time scales) and the time asymptotics are dominated by the hierarchical organization: in this regime $\alpha$ becomes the only relevant construction parameter. We remark that, being $n_i$ the number of links between any two modules of size $M_{i-1}$, the maximum value of $n_i$ is given by $M_0^2$, so that $\alpha$ can take values in the interval $4/M_0^2\leq \alpha \leq 4$.  Fig. \ref{fig:topo} shows numerical measurements of the topological dimension $D$, which is found to increase linearly with $\alpha$, in the limit of large system sizes $N$. While this result is only approximate for smaller values of $N$, where higher dimensions are complex to measure due to the size constraint, the linear behavior seems to take over for sizes around $N\approx10^6$ and above. This finding helps contextualize previous results, which pointed to a quasi-linear growth of $D$ with $\alpha$ \cite{Moretti2013}. While that result was initially dismissed as a possible effect of a too-limited $\alpha$ window and a non-exhaustive size-scaling analysis, our current results, extending up to $N=2^{24}\approx 1.7\times 10^7$, seem to robustly confirm the conjecture of a $D\sim\alpha$ dependence. While we have not been able to prove this result analytically so far, we will show in the rest of the paper its remarkable implications for activity spreading.
\section{Spectral analysis of HMNs}
The study of the epidemic threshold often relies on spectral arguments. In the case of SIS dynamics, network spectra refer to the adjacency matrix $\mathbf{A}$, whose generic element $A_{ij}$ is $1$ if nodes $i$ and $j$ are linked, and $0$ otherwise. In particular, if one considers undirected networks as we do here, $\mathbf{A}$ is symmetric and thus diagonalizable.  In most cases one is interested in the higher spectral edge of $\mathbf{A}$, including the largest eigenvalues $\Lambda_1 > \Lambda_2\ge\Lambda_3 ...$, where the uniqueness of the largest $\Lambda_1$ is ensured by the Perron-Frobenius theorem whenever the network is connected. Provided that the spectral gap $\Lambda_1-\Lambda_2$ is large, one can prove that $\lambda_c^\mathrm{QMF}= 1/\Lambda_1$, within the framework of the quenched mean-field (QMF) approximation, as we recall in the following.

The activity state for a SIS dynamic process is given by the column vector $\bm{\rho}$, whose generic element $\rho_i$ is the probability that node $i$ is in the active state at a generic time $t$, and its steady state limit $\bm{\rho^\infty}$, obtained for $t\to\infty$. Our starting point is the well-known exact result for the time evolution: $\dot{\bm{\rho}}\le -\bm{\rho}+\lambda \mathbf{A}\bm{\rho}$, which originates from the full SIS problem by neglecting (negative) quadratic terms as well as dynamic correlation contributions \cite{PastorSatorras2015}. In the $t\to\infty$ limit (when the time derivative of $\bm{\rho}$ vanishes), using the eigendecomposition $\mathrm{A}=\sum_{m=1}^N\Lambda_m \mathbf{v}_m^T\mathbf{v}_m$, where $\mathbf{v}_m$ is the $m-th$ eigenvector, we obtain   
\begin{equation}\label{eq:exact}
\bm{\rho}^\infty \leq \lambda\sum_{m=1}^{N}\Lambda_m(\mathbf{v}_m\cdot\bm{\rho}^\infty) \mathbf{v}_m, 
\end{equation} 
which expresses the steady state $\bm{\rho}^\infty$ of a SIS process of spreading rate $\lambda$ in terms of its projections on the eigenspaces of $\mathbf{A}$, with the scalar product $\mathbf{v}_m\cdot\bm{\rho}^\infty = \mathbf{v}_m^T \bm{\rho}^\infty$. The prevalence or steady-state density of active nodes is then defined as $\rho^\infty=|\bm{\rho}^\infty|$. If $\Lambda_1\gg \Lambda_2$, the sum is dominated by the leading term, the scalar product is approximately equal to unity, and a non-trivial steady state may exist only if $\lambda > 1/\Lambda_1$. In particular, the QMF result $\lambda_c^\mathrm{QMF}=1/\Lambda_1$ is recovered when the equal sign is taken in Eq. (\ref{eq:exact}).

The generality of this result in SF networks has been the focus of a recent debate \cite{Goltsev2012,Lee2013,Boguna2013}. A crucial aspect here is introduced by the localization property. An eigenvector of $\mathbf{A}$ is called localized if it has all vanishing components, except for a small subset of them. A measure of eigenvector localization is provided by the inverse participation ratio $I_m=\sum_{i=1}^N v_{mi}^4$ of eigenvector $\mathbf{v}_{m}$ \cite{Farkas2001,Goltsev2012,PastorSatorras2016}. It was initially proposed \cite{Goltsev2012,Lee2013} that in networks displaying a localized $\mathbf{v}_1$ the epidemic threshold should be replaced by an interval of the spreading rate $\lambda$, in which active states are short-lived, ideally a Griffiths phase \cite{Lee2013}, as unique unstable but localized modes do not suffice to create a global state of endemic network activity. While this view has been more recently challenged in the case of SF networks \cite{Boguna2013}, we mentioned above that HMNs are now well-known to exhibit such a Griffiths phase, whose upper bound is given by the actual critical point $\lambda_c$ \cite{Moretti2013}. Such epidemic threshold does not comply with the QMF prediction, leading to $\lambda_c>\lambda_c^\mathrm{QMF} =1/\Lambda_1$ \cite{Moretti2013}. It was argued  \cite{Moretti2013} that this result is related to the fact that in HMNs: (i) spectral gaps are small; and (ii) localization extends to all eigenvectors in the higher spectral edge of $\mathbf{A}$, that is, it is not limited to the principal eigenvector $\mathbf{v}_1$. We show here the extent to which the basic assumptions of QMF are altered in HMNs, thus making the estimate of $\lambda_c$ a formidable task. 

\begin{figure} \includegraphics[scale=.2]{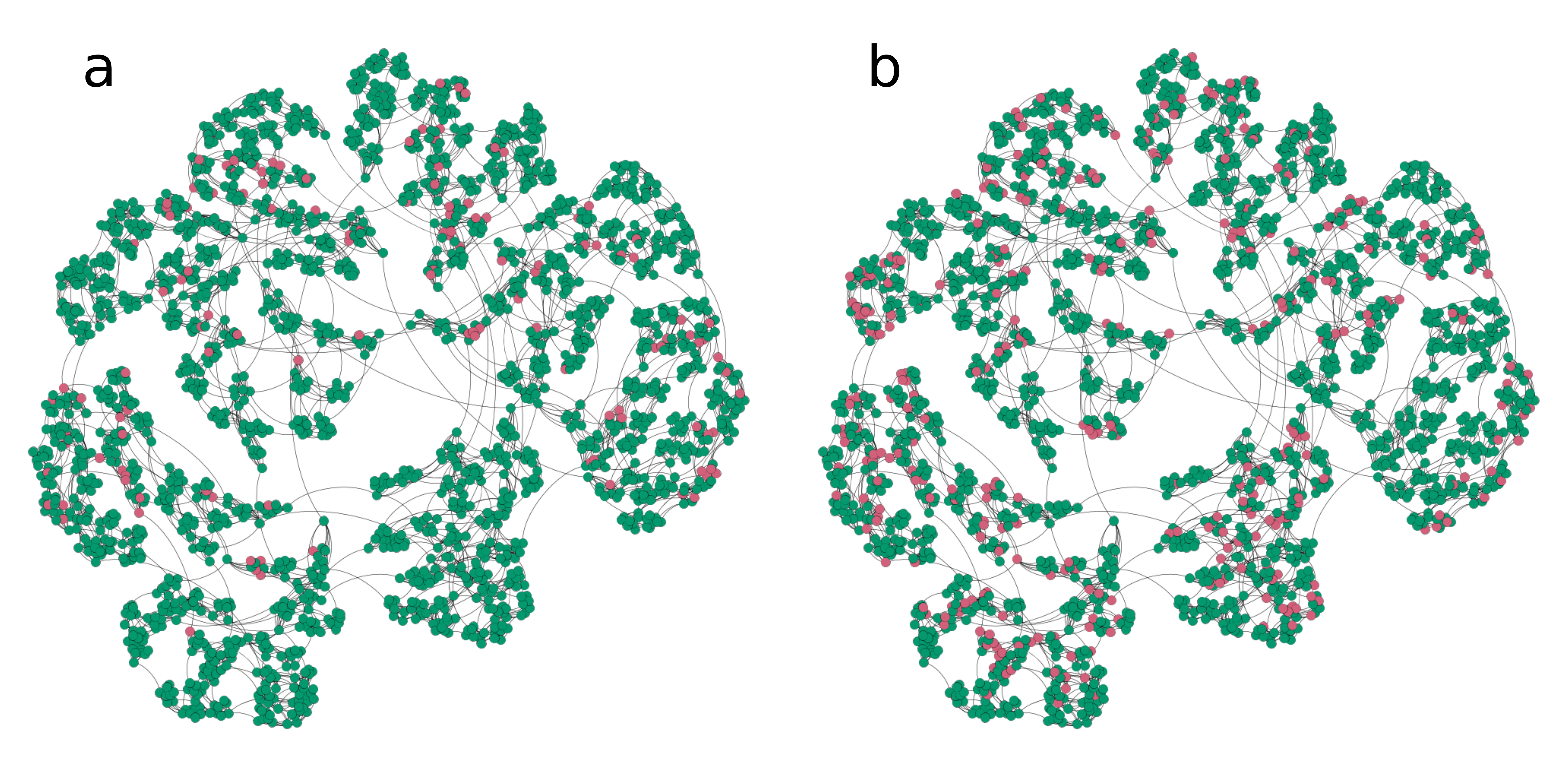} \caption{Graphical representation of the concept of dynamic coalescence in a hierarchical modular network HMN of size $N=2^{11}$, consisting in $8$ hierarchical levels, with $M_0=4$, $n_i=6$, and $\alpha=1.5$. Each figure corresponds to a snapshot of SIS dynamics, where inactive and active nodes are colored in green and red respectively. Patterns of activity are obtained through numerical simulations, as described in the main text. The network size is purposely kept small in order to make the figure clearer. Larger sizes are considered instead for the results described in the rest of the paper. (a) Activity for a metastable configuration that will eventually end in the absorbing $\rho^\infty=0$ state, obtained for a subcritical value of $\lambda<\lambda_c$; activity is localized in some regions, but cannot be sustained dynamically for long times. (b) Activity for a configuration that corresponds to a non-zero steady state $\rho^\infty$, obtained for a value of $\lambda>\lambda_c$. Here too activity appears localized and no spanning (i.e. percolating) cluster arises, however the stability of the steady state suggests that local activity patters interact \emph{dynamically} --giving rise to an effective coalescence-- resulting in a sustained global active state. }\label{fig:hmns} \end{figure}

The existence of a large spectral gap is equivalent to saying that the steady state is approximately given by the principal eigenvector $\mathbf{v}_1$: as argued above, the sum in Eq. (\ref{eq:exact}) would be dominated by the $m=1$ term and, taking the equal sign (QMF approximation), $\bm{\rho}^\infty$ would be simply given by $\mathbf{v}_1$. Such a strong statement works remarkably well in networks with strong small world properties \cite{Farkas2001} or entangled connectivity patterns \cite{Donetti2005,Donetti2006}, but fails to describe systems like ours in which spectral gaps are {\it small} \cite{Moretti2013,Villegas2014}. We propose a way to relax the above assumption, by considering a candidate steady state $\bm{\sigma}^\infty$ as a linear combination of the first $m^*$ eigenvectors, that is $\bm{\sigma}^\infty \approx \sum_{m=1}^{m^*} (\mathbf{v}_m \cdot \bm{\sigma}^\infty) \mathbf{v}_m $. In particular, we can always choose our eigenvector basis in such a way that $\mathbf{v}_m \cdot \bm{\sigma}^\infty\geq 0$ for all $m\leq m^*$. We note that in the worst case scenario $m^*=N$, meaning that all eigenvalues are needed. Under what conditions is the network able to sustain the candidate steady state, so that $\bm{\rho}^\infty=\bm{\sigma}^\infty$? By inserting our expansion for the steady state in Eq. (\ref{eq:exact}), and using the linear independence of eigenvectors, we obtain the condition
\begin{equation}\label{eq:condition}
(\lambda\Lambda_m-1)(\mathbf{v}_m \cdot \bm{\sigma}^\infty) \geq 0\;\;\; \forall m\leq m^*.
\end{equation}
 Eq. (\ref{eq:condition}) shows that in a spectral theory of SIS in which the first $m^*$ eigenvalues are necessary, also the corresponding eigenvectors play a role. Then, two mechanisms may be responsible for the lack of a non-trivial steady state with $\bm{\rho}^\infty \neq 0$:

\noindent (a) if the spreading rate is small, i.e. $\lambda<1/\Lambda_1$ (and thus, $\lambda<1/\Lambda_m$  for all $m$) then the candidate non-trivial steady state $\bm{\sigma}^\infty$ is unstable; 

\noindent (b) even if $\lambda\geq 1/\Lambda_m$ for some $m$, the localization of $\mathbf{v}_m$ induces $\mathbf{v}_m \cdot \bm{\sigma}^\infty\approx 0$ (positive but vanishing): the candidate steady state is weakly stable in principle, however dynamic fluctuations are bound to deactivate it over time.  In this case, a virtually stable state results in fact metastable and relatively short-lived \footnote{We note that a similar phenomenology has been also proposed for different types of systems in which a single eigenvalue expansion is still possible \cite{Goltsev2012}.}, in agreement with the Griffiths phase picture \cite{Moretti2013}.

Thus, a nonzero steady state is still possible, if one of the following two conditions is met:

\noindent i) there are some de-localized eigenstates, i.e. a pathological region in the spectrum exists around a given $\Lambda_m$, characterized by high density of states and/or low localization, such that it can trigger a stable $\bm{\rho}^\infty$ such that $\mathbf{v}_m \cdot \bm{\rho}^\infty \gg 0$ and $\lambda_c\approx 1/\Lambda_m$; 

\noindent ii) a finite number of localized states can \emph{dynamically coalesce} sustaining a global active state (see Fig. \ref{fig:hmns}), i.e. no pathological $(\Lambda_m,\mathbf{v}_w)$ pairs exist, however if $\lambda$ is large enough, $\lambda\Lambda_m-1>0$ for many values of $m<m^*$: a large enough number of potentially short-lived states is generated, large enough to sustain (or {\it coalesce} into) a long-lived steady state $\bm{\rho}^\infty>0$, however with no clear criterion to identify what a large enough $m^*$ is.

In order to clarify which of the two scenarios holds in the case of HMNs, and how the value of $\lambda_c$ can be related to spectral properties of $\mathbf{A}$, we conducted an extensive numerical study of HMNs of different $\alpha$, targeting both the SIS dynamics direct simulation and the network spectral properties.

Our numerical results provide a clear indication that the relevant scenario for HMNs is indeed ii) as exemplified graphically in Fig. \ref{fig:hmns}, illustrating the lack of any de-localized state \footnote{Similar results (not shown here) are obtained by looking at plots of the eigenvector inverse participation ratio $I_m$ as a measure of localization}. Similarly, we did not find any correlation between the inverse of any eigenvector and the numerically determined value of $\lambda_c$ (see below), thus indicating that if an epidemic threshold is reached --above which sustained activity exists-- it has to emerge owing to a finite number of unstable, though localized, active regions and their dynamical interplay.

\section{Scaling of the epidemic threshold in HMNs}
Given the complexity of this scenario, in which a large but undefined number of eigenvalue-eigenvector pairs is expected to play a role; is it still possible to relate $\lambda_c$ to a single scalar property of the spectrum?

\begin{figure} \includegraphics[scale=.55]{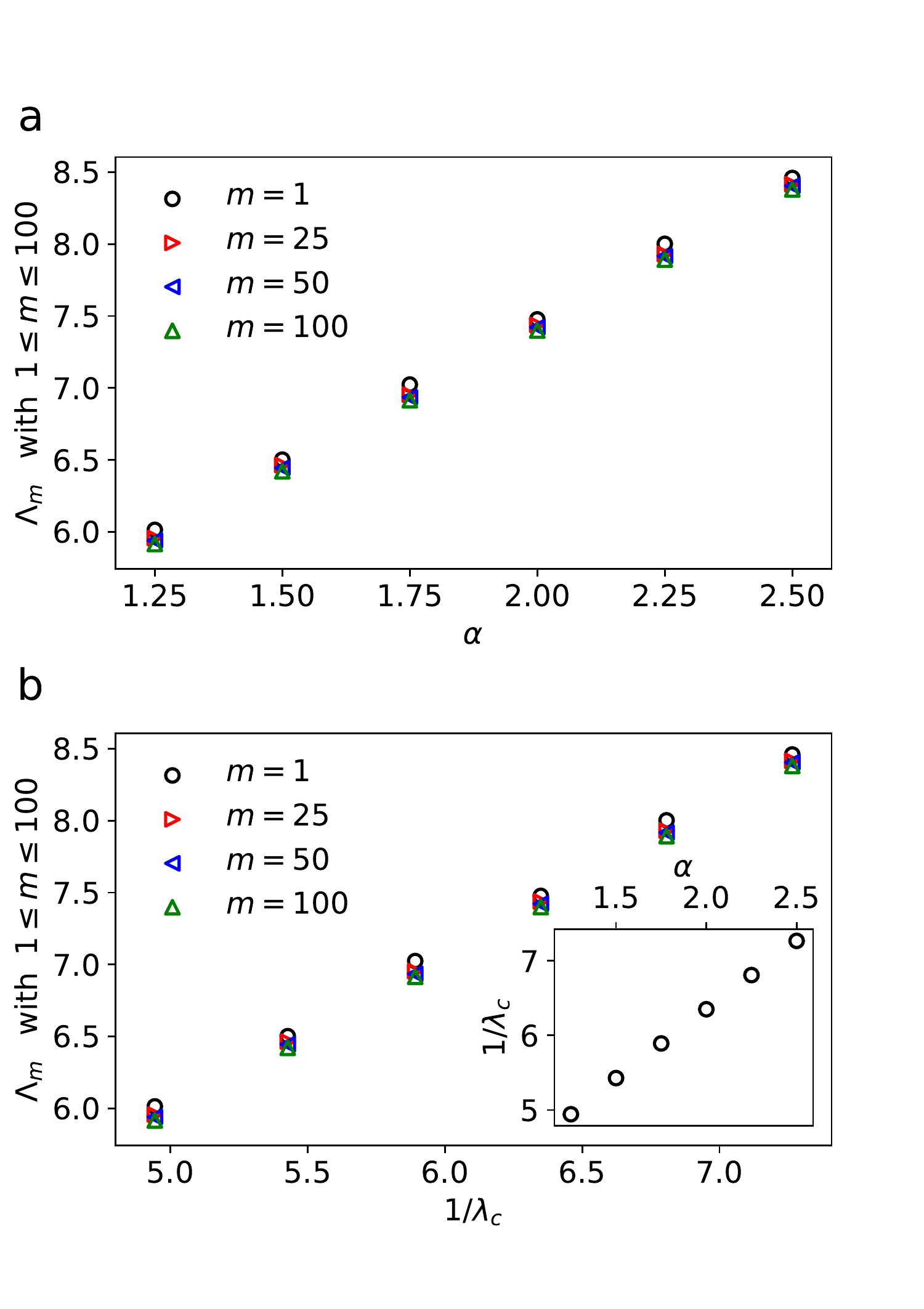} \caption{Linear dependence of {\it all} the eigenvalues $\Lambda_m$ in the spectral tail of $\mathrm{A}$ with the connectivity strength $\alpha$ (a) and the inverse of the epidemic threshold (b). While only $4$ of the $100$ largest eigenvalues are shown, this behavior extends to the whole range, and significantly beyond it. Results for networks with $N=2^{17}$ and $M_0=4$ are shown, suggesting that the behavior is robust even at moderate system sizes. The validity of the scaling relation $1/\lambda_c\sim\alpha$ is shown as en inset of plot b.}\label{fig:linear} \end{figure}

In order to answer this question, we notice that the properties of the higher spectral edge of HMNs can be tuned by acting on $\alpha$: upon increasing $\alpha$, the number of non-zero off-diagonal elements of $\mathbf{A}$ increases accordingly. A simple argument to shed light on this observation is as follows: at every hierarchical level $i$, the approximate and coarse-grained connectivity pattern between two modules would be very roughly represented by an effective weighted adjacency matrix structured as $\mathbf{A}^{(i)}\sim c_i\left( 
\begin{array}{cc}
0 & \alpha \\
\alpha & 0 \\
\end{array}
\right),$ 
which will contribute a larger eigenvalue  proportional to $c_i \alpha$ to the higher spectral edge of $\mathbf{A}$. We can corroborate our view by looking at Fig. \ref{fig:linear}a, where it is shown that the value of the generic $i$-th eigenvalue is strictly proportional to $\alpha$, for any value of $i$ in the higher spectral edge. This is unlike what happens in Erd\"os-R\'enyi (ER) or SF networks, where increasing connectivity -- for instance by tuning the average degree in ER networks or the degree distribution in SF networks -- primarily affects the scaling of the principal eigenvalue $\Lambda_1$ only, effectively changing the size of the spectral gap \cite{Farkas2001}.  We stress that our approach here is kept purposely simple, while the correct way to address the analytical computation of HMN spectra has been recently discussed for the Laplacian spectrum of the Dyson fully connected hierarchical graph \cite{Agliari2015_PRL,Agliari2016}.

The exceptional behavior of HMN spectra, for which a single tuning parameter is able to tune the entire spectral edge, turns crucial for our study of the epidemic threshold:
provided that $\lambda_c$ scales as the inverse of some relevant eigenvalues in the higher spectral edge, the similarity properties of eigenvalue spectra $\Lambda_m\sim \alpha$ provides us with a simple prediction for the epidemic threshold, which can be expressed as:
\begin{equation}\label{eq:prediction}
\lambda_c \sim \frac{1}{\alpha}
\end{equation}

We measured $\lambda_c$ computationally by employing the standard method of simulating SIS dynamics starting from a homogeneous initial condition of all active nodes, and recording steady states values (more refined techniques relying on susceptibility measurements have been proposed for SF networks, in which vanishing values of $\lambda_c$ make their estimate a more delicate task \cite{Ferreira2012}).  Simulations we performed for sizes $N=2^{17}$ and $N=2^{20}$, recording minimal to no deviations between the two cases. From the perspective of size scaling, it appears that in this size range ($10^5$ -- $10^6$ nodes) the dynamics display no appreciable size effects. In particular, we find that steady states for $\lambda>\lambda_c$ are stable against fluctuations, and do not decay as a consequence of finite sizes. Smaller sizes, however, may still allow large enough fluctuations to make steady states unstable. This apparent inconsistency (an unclear distinction between supercritical and subcritical dynamics), affects systems of very limited sizes, and is of course understandable considering that phase transitions are correctly defined only in the $N\to\infty$ limit.  The resulting $\lambda_c$ is plotted in the inset of Fig. \ref{fig:linear}b as a function of $\alpha$.

The validity of Eq. (\ref{eq:prediction}) is confirmed by simulation results in Fig. \ref{fig:linear}b, where it is shown that there is a remarkable scaling property \begin{equation} \Lambda_m\sim \frac{1}{\lambda_c} \end{equation} for all eigenvalues in the spectral edge. In passing, we note that as all eigenvalues in the range scale with $1/\lambda_c$, this must happen, in particular, for $\Lambda_1$: while the principal eigenvalue alone by itself cannot justify the value of the epidemic threshold, the standard $\lambda^\mathrm{QMF}_c=1/\Lambda_1$ criterion survives in a weaker form as a scaling law (not an equality), and more importantly, it extends to all eigenvalues that participate in the onset of activity.  

Finally, recalling our initial results, that for large enough networks one has $D\sim\alpha$, we can rewrite our estimate as 
\begin{equation}
\lambda_c\sim \frac{1}{D}.
\end{equation}  
The conjecture that the network dimension acts as a structural determinant of spreading activity appears corroborated: $D$ is indeed capable of tuning the value of $\lambda_c$.  The picture emerging from our results is that of a complex dynamic scenario, as the one described by Eq. (\ref{eq:condition}), which can be correctly understood by identifying the topological dimension $D$ as the relevant tuning parameter for epidemic spreading in HMNs.  

We also remark that in the $D\to\infty$ limit, Eq. (\ref{eq:prediction}) predicts a vanishing epidemic threshold, recovering the well-known quenched mean field result for SF networks, which are intrinsically infinite dimensional. The actual study of the existence of a finite $\lambda_c$ in SF graphs requires in fact more refined theoretical tools, while here we only stress how our generalization of the QMF approach also contains its original predictions.

\begin{figure} \includegraphics[scale=.55]{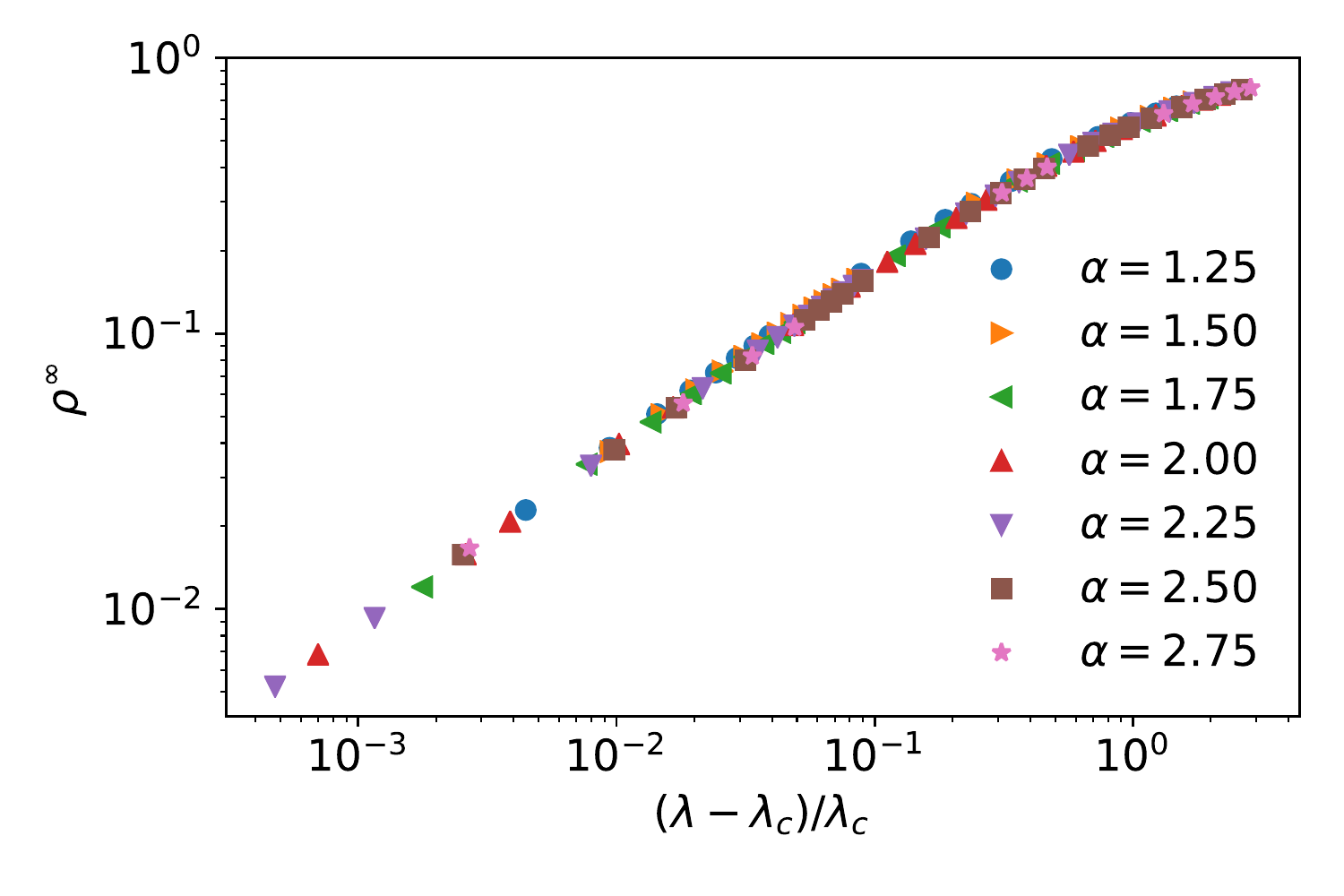} \caption{Dependence of the steady state values of the density of active sites (our order parameter) on the relative distance from the critical spreading rate (our control parameter), as obtained in SIS dynamics simulation on HMNs with $N=2^{20}$ and $M_0=4$, for different values of $\alpha$. All curves collapse, pointing to scale invariant behavior, with a non trivial exponent $\beta\approx 0.65$.}\label{fig:steady} \end{figure}

Finally, we addressed the question of how the supercritical phase diagram is affected by tuning $D$, or equivalently $\alpha$. Fig. \ref{fig:steady} shows the steady state density of active nodes, $\rho^\infty$, as a function of $(\lambda-\lambda_c)/\lambda_c$, for different values of $\alpha$ (and $D$). The striking result that we find is that while the emergence of scaling clearly suggests a critical scenario of the familiar form $\rho^\infty\sim (\lambda-\lambda_c)^\beta$, even for the lower values of $\alpha$ considered here the system is above its critical dimension and the $\beta$ exponent is insensitive to dimensionality. A detailed study of the phase transition for even lower dimensional HMNs is beyond the scope of this manuscript, and is being considered for future work. Here, we would like to emphasize that the emergence of dynamic scaling is robust against structural variations, in a range of connection densities and dimensions which we consider relevant in fields such as neuroscience.

\section{Conclusions}

We discussed the essential reasons why the QMF approximation does not provide a correct prediction for the epidemic threshold in hierarchical modular networks. The first eigen-mode to become unstable is localized and thus, it cannot represent an endemic state 
of activity once fluctuations are taken into consideration. Instead, a finite number of unstable eigen-modes may induce an active state.
Remarkably such set of eigenvectors does not create a spanning or percolating cluster expanding through the whole network, but just localized clusters of activity. However such ``reservoirs'' of instability can create a global active state owing to fluctuations that transiently activate other modes, effectively connecting diverse localized clusters.

 Consequently, we introduced a novel framework, in which the epidemic threshold depends on a (large) number of eigenvalues but, remarkably, turns out to be inversely proportional to a unique parameter; the network topological dimension, $D$. While our results corroborate this result for HMNs, the possibility of extending it to more general network classes can only be conjectured at this stage.

The importance of our results in the description of biological systems endowed with hierarchical modular organization can be better understood considering the example of brain networks. A direct dependence of $\lambda_c$ on $D$ ensures that, being HMNs finite dimensional, the epidemic threshold never vanishes. Should that happen, information units would boundlessly propagate through the network, making it impossible to manage information properly, something usually associated with epileptic forms of activity. On the other hand, by simply tuning a single parameter, e.g. during the development or pruning of neuronal networks, the dynamical regime
of the whole network could be established.

In conclusion, we have studied numerically some of the most relevant properties of the onset of spreading in HMNs. We have highlighted a crucial construction parameter, the connectivity strength $\alpha$ --that controls the network fractal-like structure and shown its proportionality to the topological dimension of a HMN. We have shown how $\alpha$ and, equivalently-- $D$ are responsible for the tuning of the epidemic threshold $\lambda_c$ in HMNs, by acting directly on the spectral properties of the adjacency matrix.  Thus, by slightly modifying a unique network-structure controlling parameter, it is possible to regulate the spreading rate that is necessary to generate sustained activity, i.e. the ``epidemic threshold'' and, in this way, the overall state of activity can be controlled by the network architecture.

We hope that our work can stimulate further studies in this direction, possibly providing a deeper analytical understanding of the relationship and interplay between structural and dynamic patterns of localization.

\begin{acknowledgments} 
  AS and PM acknowledge financial support from the Deutsche Forschungsgemeinschaft, under grant MO 3049/1-1.  MAM is grateful to the Spanish-MINECO for financial support (under grant FIS2013-43201-P; FEDER funds).  We thank P. Villegas and S. di Santo for useful discussions and for a critical reading of the manuscript. \end{acknowledgments}

%\vspace{-0.65cm}
%\bibliographystyle{prsty}
%\def\url#1{}
%\bibliography{Biblio}

\end{document}